\documentclass[10pt,conference]{IEEEtran}
\usepackage{mathpazo}
\usepackage{times}

\usepackage{amsmath}
\usepackage{amsfonts}
\usepackage{latexsym}
\usepackage{amssymb}

\usepackage{upref}
\usepackage{theorem}
\usepackage{graphicx}
\usepackage{psfrag}
\usepackage{algorithmic}
\usepackage{algorithm}
\usepackage{verbatim}
\usepackage{cite}
\usepackage{color}

\theoremstyle{plain}
\theorembodyfont{\normalfont\slshape}

\newtheorem{thm}{Theorem$\!$}

\newtheorem{lem}[thm]{Lemma$\!$}

\newtheorem{claimm}[thm]{Claim$\!$}

\newtheorem{prop}[thm]{Proposition$\!$}

\newtheorem{cor}[thm]{Corollary$\!$}

\newtheorem{defn}[thm]{Definition$\!$}

\newtheorem{xmpl}[thm]{Example$\!$}

\newtheorem{cnstr}{Construction$\!$}

\def\qed{\hskip 3pt \hbox{\vrule width4pt depth0pt height4pt}}

\newcounter{enumrom}
\renewcommand{\theenumrom}{(\roman{enumrom})}

\makeatletter
\renewcommand{\@endtheorem}{\endtrivlist}
\makeatother

\makeatletter
\renewcommand{\thefigure}{{\@arabic\c@figure}}
\renewcommand{\fnum@figure}{{\bf Figure\,\thefigure}}
\makeatother

\newcommand{\mira}[1]{[{\bf Mira's remark}: {#1}]}

\newcommand{\suppress}[1]{}

\def\01{\{0,1\}}

\newcommand{\remove}[1]{}

\newtheorem{theo}{Theorem}
\newtheorem{clm}{Claim}

\newcommand{\BA}{\begin{alg}} \newcommand{\EA}{\end{alg}}
\newcommand{\BE}{\begin{enumerate}} \newcommand{\EE}{\end{enumerate}}
\newcommand{\BT}{\begin{theo}} \newcommand{\ET}{\end{theo}}
\newcommand{\BL}{\begin{lem}} \newcommand{\EL}{\end{lem}}
\newcommand{\BCM}{\begin{clm}} \newcommand{\ECM}{\end{clm}}
\newcommand{\BI}{\begin{itemize}} \newcommand{\EI}{\end{itemize}}

\newenvironment{prf}{\noindent{\bf Proof:~~}}{\(\qed\)}
\newcommand{\BPF}{\begin{prf}} \newcommand {\EPF}{\end{prf}}
\newenvironment{proofof}[1]{\noindent{\bf Proof of {#1}.~}}{\endprf}
\newcommand{\BPFOF}{\begin{proofof}} \newcommand {\EPFOF}{\end{proofof}}

\newcommand{\BEQN}{\begin{eqnarray}}\newcommand{\EEQN}{\end{eqnarray}}
\newcommand{\BEQ}{\begin{equation}} \newcommand{\EEQ}{\end{equation}}

\newcommand{\eat}[1]{}

\begin{document}

\IEEEoverridecommandlockouts

\title{Coded Cooperative Data Exchange Problem for General Topologies}

\date{}

\author{
\IEEEauthorblockN{\textbf{Mira Gonen}}
\IEEEauthorblockA{Department of Mathematics  \\
Bar-Ilan University \\
{\it gonenm1@math.biu.ac.il }\vspace*{-4.0ex}}
\and
\IEEEauthorblockN{\textbf{Michael Langberg}}
\IEEEauthorblockA{Department of Mathematics and Computer Science\\
The Open University of Israel \\
{\it mikel@open.ac.il}\vspace*{-4.0ex}}
\thanks{
  This work was supported in part by ISF grant 480/08 and
  the Open University of Israel's research fund
  (grant no.~46114). Work done in part while Mira Gonen was at the Open University of Israel.}
}

\date{}
\maketitle
\begin{abstract}
We consider the {\em coded cooperative data exchange problem} for general graphs.
In this problem, given a graph $G=(V,E)$ representing clients in a broadcast network, each of which initially hold a (not necessarily disjoint) set of information packets; one wishes to design a communication scheme in which eventually all clients will hold all the packets of the network.
Communication is performed in rounds, where in each round a single client broadcasts a single (possibly encoded) information packet to its neighbors in $G$.
The objective is to design a broadcast scheme that satisfies all clients with the minimum number of broadcast rounds.

The coded cooperative data exchange problem has seen significant research over the last few years; mostly when the graph $G$ is the complete broadcast graph in which each client is adjacent to all other clients in the network, but also on general topologies, both in the fractional and integral setting.
In this work we focus on the integral setting in general undirected topologies $G$.
We tie the data exchange problem on $G$ to certain well studied combinatorial properties of $G$ and in such show that solving the problem exactly or even approximately within a multiplicative factor of $\log{|V|}$ is intractable (i.e., NP-Hard).
We then turn to study efficient data exchange schemes yielding a number of communication rounds comparable to our intractability result.
Our communication schemes do not involve encoding, and in such yield bounds on the {\em coding advantage} in the setting at hand.
\end{abstract}

\section{Introduction}


In this work we study the coded cooperative data exchange problem for general graphs.
An instance to the problem consists of an undirected graph $G=(V,E)$ representing a communication network (in which each node of $G$ represents a client, and edges in $G$ represent client pairs that can communicate with each other), a parameter $k$ representing the number of information packets $X=\{x_1,\dots,x_k\}$ to be transmitted over the network, and a set $\{X_i\}_{i \in V}$ of subsets of $X$ representing the set of packets available at each client node $v_i \in V$ in the initial stage of the transition.
The objective is to design a communication scheme in which, eventually, all nodes of the network will obtain all $k$ packets.
Loosely speaking, in each round of the communication scheme, a single node broadcasts a single (possibly encoded) packet to all its neighbors in $G$.
The goal is to find a communication scheme in which the number of communication rounds is minimum.

The coded cooperative data exchange problem has seen significant research over the last few years.
The problem was introduced by El Rouayheb et al. in \cite{RSS10}, where data exchange over a {\em complete} graph $G$ was considered (in which each client can broadcast its messages to all other clients in $G$). In \cite{RSS10} certain upper and lower bounds on the optimal number of transmissions needed was established.
In a subsequent work, Sprinston et al. \cite{SSBR10} continue the study of complete graphs $G$ and present a (randomized) algorithm that with high probability
achieves the minimum number of transmissions, given that the packets are elements in a field $F_q$ with $q$ large enough.
Ozgul et al. \cite{OS11} study a variant of the data exchange problem in which each client has a distinct broadcast cost and one wishes to minimize the cost of the transmission scheme after which all clients have obtained all information packets. In \cite{OS11}, optimal randomized linear encoding schemes are given for the problem at hand.

Communication in which {\em fractional} packets can be transmitted is addressed in the works of Courtade et al. in \cite{CXW10} (for general topologies $G$) and Tajbakhsh et al. \cite{TSS,TSS11} (for the complete topology).
In the fractional setting, packets are assumed to be divisible into chunks so that a fraction of a packet may be transmitted at any (fractional) round of communication; as apposed to the integral setting in which information packets are indivisible.
In \cite{CXW10,TSS,TSS11} it is shown that the fractional setting of the data exchange problem reduces to that of multicast network coding and can be efficiently solved in an optimal manner via linear programming and the concept of linear network coding, see e.g. \cite{ACLY00,LYC03,KM03,JSCEEJT04,HoMKKESL06}.

Most related to our work is the work of Courtade et al. in \cite{CW10} which focus on general topologies $G$ in the setting of indivisible packets (the integral setting). \cite{CW10} continue the paradigm of \cite{CXW10} which characterizes the data exchange problem as a family of {\em cut inequalities}, and present certain communication schemes that yield approximate solutions for an asymptotic number of packets $k$.
Roughly speaking, \cite{CW10} analyze a certain communication scheme in which each client transmits at a certain {\em fixed rate} over time, and obtain nearly optimal rate allocations (within an additive approximation of $\varepsilon k$ for general graphs, and $|V|$ for regular graphs).
An important aspect in the analysis in \cite{CW10} is the assumption that the number of packets $k$ tends to infinity.
A detailed comparison of the results of \cite{CW10} with ours appears below at the end of Section~\ref{sec:oc}.

Most recently, Milosavljevic et al. \cite{MPRGR11} present a comprehensive study of data exchange over the complete topology in which one wishes to broadcast the components of a (jointly distributed) discrete memoryless multiple source. Efficient optimal {\em rate} schemes are presented for a number of side information models.

\subsection{Our contribution}
\label{sec:oc}
In this work we study the coded cooperative data exchange problem on general topologies.
We focus on the combinatorial {\em integral} setting in which one assumes that packets are indivisible.
Namely, we assume that each packet is a value from a given alphabet $\Sigma$, and in each communication round a single element of $\Sigma$ is broadcasted by a client to its neighbors in $G$.
The study of the indivisible integral setting, rises naturally in communication schemes in which dividing information packets to several chunks leads to undesirable overhead in communication (via scheduling issues or rate loss due to header information).
Our work addresses the design and analysis of {\em efficient} algorithms that (approximately) solve the problem at hand.
Throughout our work, we assume that the number of packets $k$ is polynomial in the size of the network $|V|$. In this context, an efficient algorithm is one which is polynomial in the network size.

We start by tying the data exchange problem in general topologies $G$ to certain well studied combinatorial properties of $G$.
Specifically, we consider the {\em Dominating Set} problem (e.g., \cite{j74}) and its variants (to be defined in Section~\ref{sec:defns}), and show that they are closely related to the data exchange problem.
Namely, we show that (i) a solution to the Dominating Set problem (or its variants) yields a (not necessarily optimal) solution to the data exchange problem, and (ii) an optimal solution to the data exchange problem yields a nearly optimal solution to the Dominating Set problem(s).
Roughly speaking, these connections (together with others) imply two initial results.
Primarily, that it is NP-Hard to find a solution to the data exchange problem in which the number of communication rounds is within a multiplicative factor of $\Omega(\log \vert V\vert)$ from the optimal.
Secondly, that a conceptually simple data exchange algorithm, that does not involve encoding, based on the Dominating Set problem yields a number of communication rounds which is within a multiplicative factor of $O(k \cdot \log \vert V\vert)$ from the optimal.

The gap between the upper and lower bounds above is $k$ (the number of distinct packets in the network) which may be of significant size. Reducing this gap is the main focus of our work.
Roughly speaking, in this work we reduce the gap of $k$ by analyzing our algorithm based on the Dominating Set problem(s).
Our algorithm does not involve coding and in such yields bounds on the {\em coding advantage} in the setting of data exchange.
Our detailed results are given below, which at times are the best possible (assuming standard tractability assumptions).

The paper is structured as follows.
In Section~\ref{sec:defns}, we present the model and notation used throughout this work, including the several variants of the Dominating Set problem used in our analysis.
In Section~\ref{sec:intract}, we prove that it is NP-Hard to approximate the data exchange problem on general topologies within a multiplicative factor of $\Omega(\log{n})$ (for any $k$ polynomial in $n$).
Here, $n=|V|$.

In Section~\ref{sec:alg}, we present our algorithm for data exchange based on the Dominating Set problem and its variants.
The algorithm we present is conceptually very simple and does not involve coding.
As mentioned above, a naive analysis of our algorithm yields an approximation ratio of $O(k \cdot \log n)$, and the majority of this section is devoted to proving that the algorithm actually performs better.

In Section~\ref{sec:disjoint}, we show that our algorithm is the best possible (assuming standard tractability) and has an approximation ratio of $O(\log{n})$ (matching the lower bound of Section~\ref{sec:intract}) on instances in which each packet is initially present at a single client in $G$.
This implies a coding advantage of $O(\log{n})$ in such cases.

In Section~\ref{sec:regular}, we study data exchange instances in which the underlying graph is regular (each client has the same number of neighbors). We show that the approximation ratio in this case is again better than $O(k \cdot \log \vert V\vert)$ and depends on the average number $\bar{d}$ of packets available at client nodes.
Specifically we show that in this case the approximation ratio of our algorithm is
$O\left(\frac{k}{k-\bar{d}}\right)\log{n}=O\left(1+\frac{\bar{d}}{k-\bar{d}}\right) \log{n}$ (thus improving the factor of $k$ in the naive analysis to $1+\frac{\bar{d}}{k-\bar{d}}$). Notice, that for $\bar{d} = \Theta(k)$ (the case in which on average each client initially has a constant fraction of the packets) we obtain an approximation ratio that matches the bound of Section~\ref{sec:intract}.
Our results imply a coding advantage of $O\left(1+\frac{\bar{d}}{k-\bar{d}}\right) \log{n}$ in the cases at hand.
Finally, in Section~\ref{sec:General Case} we study general graphs $G$ with no restrictions and present an improved approximation ratio to that naively mentioned above.

We conclude our work by studying a refined version of our algorithm (still without encoding) in Section~\ref{sec:talg} and by discussing future research directions in Section~\ref{sec:diss}.

Comparing our results with those in \cite{CW10} is not straightforward.
Courtade et al. \cite{CW10} focus on the setting in which the number of packets $k$ tends to infinity and may be significantly greater than the network size $n$.
The setting of asymptotic $k$ allows the design of algorithms which are efficient with respect to $k$ but may be exponential in $n$.
In our work we focus on the setting in which $k$ is polynomially bounded by $n$, and obtain communication schemes that can be designed efficiently in time polynomial in the network size $n$.
In addition, \cite{CW10} focus on the case in which every client initially holds a constant fraction of the $k$ information packets;\footnote{The precise formulation in \cite{CW10} is phrased in terms of ``well behaved'' packet distributions; i.e., the asymptotic (in $k$) empirical probability that a client (or set of clients) holds a certain number of packets.} and in this setting
study additive approximations.
In this work, we study multiplicative approximations, and our assumptions (if any) on the packet distribution are of different nature.


%
%

\section{Model Definition and Preliminaries}
\label{sec:defns}


\subsection{Coded Cooperative Data Exchange Problem}
We start by defining the Coded Cooperative Data Exchange Problem for General Graphs.
Let $G=(V,E)$ be a given undirected graph with $V = \{v_1, . . . , v_n\}$.
Let $X = \{x_1, . . . ,x_k\}$ be a set of packets to be delivered to the $n$ clients belonging to the set
$V$. The packets are elements of a finite
alphabet which will be assumed to be a finite field
$F_q$. At the beginning, each client $v_i$
knows a subset of packets denoted by $X_i\subseteq X$, while
the clients collectively know all packets in $X$.
We denote by $\bar {X_i} = X \setminus X_i$ the set
of packets required by client $v_i$.
For each client (vertex in $G$) $v_i$ let $d_{v_i}=|X_i|$ be the number of packets it holds, let $\bar d=\sum_{v\in V}d_v/n$ be the average number of packets present at vertices of $G$, and let $d=\max_{v\in V}d_v$ be the maximum number if packets that any client holds. We will use these parameters in our analysis.

Each client may transmit packets to it neighbors in $G$ via a lossless broadcast
channel capable of transmitting a single element in $F_q$.
The data is transmitted in
communication rounds, such that at round $i$ one of the
clients, say $v$, broadcasts an element $x \in F_q$ to all its neighbors in $G$. The transmitted information $x$ may be one of the
original packets in $X_j$, or some encoding of packets in $X_j$ and
the information 
previously transmitted to $v_j$.

Our goal is to devise a scheme that enables
each client $v_i\in V$  to obtain all packets in $\bar{X_i}$ (and thus in $X$) while
minimizing the total number of broadcasts.
This work focuses on the {\em integral} (i.e., {\em scalar}) setting in which each broadcast consists of a single element of $F_q$.
We denote by $NC$ the minimum number of (integral) broadcasts needed to satisfy the given instance to the coded cooperative data exchange problem at hand.
In this work we connect the value of $NC$ with other well studies combinatorial operators on $G$ defied below.

Throughout our work, we assume that the number of packets $k$ is polynomial in the size of the network $|V|$ (i.e., $k \leq |V|^c$ for some constant $c$). In this context, we say an algorithm is  {\em efficient} if its running time is polynomial in the network size.

\subsection{The Self Dominating Set problem}

Given an undirected graph $G=(V,E)$, a {\em self dominating set} of $G$ is a subset of vertices $S$ such that every $v \in V$ is connected to some vertex $s \in S$ by an edge $(s,v) \in E$.
In such a case we say that $v \in N(s)$ where $N(s) = \{v \mid (s,v) \in E\}$.
The self dominating set problem is closely related to the standard dominating set problem, e.g. \cite{j74}, on which we elaborate below.
The minimum size of a self dominating set in $G$ is denoted by $DS^+$. 
A self dominating set $S$ with a corresponding induced subgraph that is connected is referred to as a connected self dominating set.
Denote by $CDS^+$ the size of a minimum connected self dominating set in $G$. 
We will show below that computing (or approximating) any of the values mentioned above (i.e., $DS^+,CDS^+$)
is NP-Hard.

In this work we will also be interested in a fractional version of the Self Dominating Set problems expressed by the following linear program.
Given a graph $G=(V,E)$, find a set of capacities $C=\{c_v|v\in V\}$ (where for each $v\in V$, $c_v$ is the capacity of vertex $v$) such that $\sum_{v \in V} c_v$ is minimum, and  $\forall v \in V$ it holds that $\sum_{u\in {N(v)}} c_u \geq 1$.
The above is equivalent to the solution of the following LP:

\medskip

$\begin{array}{ll}$
               Minimize $&\sum_{v \in V} c_v \\$
               subject to $&\sum_{u\in {N(v)}} c_u \geq 1$, $\forall v \in V \\
               & 0\le c_v\le 1$, $\forall v \in V
          \end{array}$
\medskip

Let $DS^+_f$ denote the minimum value of the linear program above.
By considering integral values of $c_v$, it is straightforward to establish that $DS^+_f \leq DS^+$.

%
%

As we will see, at times we would like to ``cover'' each vertex in $G$ more than once by our self dominating sets $S$.
We thus consider the integer and fractional $k$ Self Dominating Set problems as well.
Below we phrase the fractional version, with optimum denoted by $(k-DS^+)_f$, the integer variant is obtained by setting $c_v \in \{0,1\}$ and its optimum will be denoted by $k-DS^+$:
\medskip

$\begin{array}{ll}$
               Minimize $&\sum_{v \in V} c_v \\$
               subject to $&\sum_{u\in {N(v)}} c_u \geq k$, $\forall v \in V \\
               & 0\le c_v\le 1$, $\forall v \in V
          \end{array}$
\medskip

Finally, as we will see, to connect the cooperative data exchange problem with the notion of dominating sets in $G$, we will need to specify the ``cover'' requirement explicitly for each vertex $v$.
We refer to this variant as the {\em Augmented-$k$-Fractional Self Dominating Set} problem.
Here, we solve the same linear program with the exception that each vertex needs to be covered at least $k-d_v$ times (the use of the parameter $d_v$ that was defined previously to be the number of initial packets present at $v$ is not occasional).

\medskip
$\begin{array}{ll}$
               Minimize $&\sum_{v \in V} c_v \\$
               subject to $&\sum_{u\in {N(v)}} c_u \geq k-d_v$, $\forall v \in V \\
               & 0\le c_v\le 1$, $\forall v \in V
          \end{array}$
\medskip

We denote by $A-(k-DS^+)_f$ the optimal solution to the linear program above.
Note that the above is an augmented version of the $k$ fractional self dominating set problem when there is an initial
solution $\{d_v\}$ and we wish to {\em augment} it to a full solution by using
values of $\{c_v\}$.

Some observations and related work expressing the relationships between the notions defined above are in place:

\BL\label{lem:kds}
$(k-DS^+)_f=k\cdot DS^+_f.$
\EL
\BPF
Any solution $\{c_v\}$ to $DS^+_f$ implies a solution $\{c_v^k\}=\{k\cdot c_v\}$ to $(k-DS^+)_f$ and visa versa.
\EPF

Note that the above lemma is not valid for the integral versions of the problems, namely $k-DS^+\neq k\cdot DS^+.$
E.g., it is not hard to verify that the $2$ by $3$ complete bipartite graph ($K_{2,3}$) with an additional edge between the two vertices in the 2-size side has $2-DS^+=3$ and $DS^+=2$.

\BL\label{lem:augmented}
Defining the parameters $d_v$ to be equal to $|X_i|$ for every $v_i \in V$, it holds that $A-(k-DS^+)_f\le NC$.
\EL
\BPF
Consider any solution to the coded cooperative data exchange problem.
For every vertex $v \in V$, let $c_v$ be the number of times $v$ transmitted information during the execution of the solution at hand.
By our definitions $\sum c_v \geq NC$.
We now show that $\{c_v\}$ is also a solution to $A-(k-DS^+)_f$.
Namely, consider any $v \in V$ that is missing $k-d_v$ packets in our data exchange problem.
It must be the case, that during the process of communication it received at least $k-d_v$ broadcasts, as otherwise it could not be able to obtain all $k$ packets after the communication process.
Thus it holds that $\sum_{u\in {N(v)}} c_u \geq k-d_v$ as desired.
\EPF

\BL\label{lem:augmented-lb}
Let $\{d_v\}$ be the set of weights in the augmented $k$-dominating set problem, and let $d=\max_{v\in V}{d_v}$. Then
$$
(k-d)\cdot DS^+_f\le A-(k-DS^+)_f \le NC.
$$
\EL
\BPF
The right inequality follows from Lemma~\ref{lem:augmented}.
For the left inequality, we notice that each solution to the fractional augmented $k$ self dominating set problem is a fractional solution to the $(k-d)$ self dominating set problem .
Namely, let $\{c_v\}$ be the capacities of an optimal solution to the fractional augmented $k$ self dominating set problem.
Then for all $v$ it holds that
$\sum_{u\in N(v)}{c_u}\ge k-d_v\ge k-d$. Therefore, $\{c_v\}$ is a solution to the fractional $(k-d)$ self dominating set problem.
Now using Lemma~\ref{lem:kds}, we obtain:
$$
(k-d)\cdot DS^+_f = ((k-d) - DS^+)_f \le A-(k-DS^+)_f.
$$
\EPF

\subsection{The (standard) dominating set problem}
We now address the standard dominating set problem, which slightly differs from the previously defined {\em self} dominating set problem.
Given an undirected graph $G=(V,E)$, a (standard) dominating set of $G$ is a subset of vertices $S$ such that every $v \in V$ is either in $S$ or connected to some vertex $s \in S$ by an edge $(s,v) \in E$.
The minimum sized dominating set in $G$ is denoted by $DS$.
The fractional variant of the dominating set problem is expressed by the following linear problem:

\medskip
$\begin{array}{ll}$
               Minimize $&\sum_{v \in V} c_v \\$
               subject to $&\sum_{u\in {N(v)}\cup \{v\}} c_u \geq 1$, $\forall v \in V \\
               & 0\le c_v\le 1$, $\forall v \in V
          \end{array}$
\medskip

We denote the optimal solution to the linear problem above by $DS_f$
Clearly, it holds that $DS \leq DS^+$ and that $DS_f \leq DS^+_f$.

As before, one can define the connected variant of the dominating set problem, and the $k$-dominating set problem.
We denote the optional values in these cases as $CDS$ for the connected variant, $k-DS$ for integral $k$-dominating set, and $(k-DS)_f$ for fractional $k$-dominating set.
As in Lemma~\ref{lem:kds} we have that:
\BL\label{lem:kds+}
$(k-DS)_f=k\cdot DS_f.$
\EL

The following lemma that constructively connects between dominating sets and their connected variant was proven in \cite{gk98}.

\BL[\hspace{-0.01mm}\cite{gk98}]
\label{lem:cds}
Given any dominating set $D$, one can efficiently construct a {\em connected} dominating set $D'$ with $|D'| \leq \frac{4}{3}\cdot|D|$. 
Specifically, for every connected graph $G=(V,E)$ it holds that $CDS \le \frac{4}{3}\cdot DS$.
\EL


It is NP-Hard to estimate the size of the minimum dominating set of a given graph $G$ up to a multiplicative factor of $\Omega(\log{|V|})$ \cite{rs97}.
Notice that if $CDS > 1$, then $CDS^+=CDS$, (and in general $CDS^+ \leq CDS+1$) so finding $CDS$, and $CDS^+$ (and also approximating them beyond a ratio of $\Omega(\log{|V|})$) is also NP-hard. Lemma~\ref{lem:cds} and the definition of the self dominating set problem imply the following lemma which connects $DS,\ DS^+,\ CDS$, and $CDS^+$:
\BL\label{lem:CDS-DS^+}
$$
\frac{4}{3}DS+1 \ge CDS + 1\ge CDS^+ \ge DS^+\ge DS.
$$
\EL
Lemma~\ref{lem:CDS-DS^+} implies that all the values $DS,\ DS^+,\ CDS,$ and $CDS^+$ are all all approximately (up to constant factors) the same size.

\section{Intractability Results}
\label{sec:intract}
In this section we show that the coded cooperative data exchange problem is hard to approximate within a multiplicative factor of $c\log \vert V\vert$, for some $c > 0$, for every value of $k$.
We use the fact that it is NP-hard to estimate $DS$ within a multiplicative factor of $c\log \vert V\vert$, for some $c > 0$ \cite{rs97}.
We first show our hardness for $k=1$.
We then turn to the case of general $k$ (polynomial in $n$).

\BL
\label{lem:k1}
The coded cooperative data exchange problem with $k=1$ is NP-hard to approximate within $c\log \vert V\vert$, for some $c > 0$.
\EL
\BPF
We show that, essentially, the coded cooperative data exchange problem when $k=1$ is equivalent to the connected dominating set problem.
Namely, consider any (connected) instance $G=(V,E)$ of the dominating set problem and construct an instance to the data exchange problem which includes the network $G$ and a single node $v_0 \in V$ that holds the (single) message $x_1$.
We show that the number of rounds in the optimal solution to the data exchange instance at hand $NC$  is approximately the size of the minimum connected dominating set size $CDS$ of $G$.
Specifically
$$
CDS \leq NC \leq CDS+1
$$

Consider an optimal solution to the data exchange problem.
Notice that, as each edge has unit capacity, once there is only a single message $x_1$ to be broadcasted throughout the network, no encoding is needed.
Thus, any solution to the data exchange problem will correspond to a series of broadcasts of message $x_1$ at certain nodes of the network.
As there is only a single message, it also holds that no vertex needs to broadcast more than once.
Let $S$ be the set of vertices that performed a broadcast.
The size of $S$ is exactly the value of $NC$ on the instance at hand.
In addition, as every vertex $v \in V$ has received $x_1$, it holds that either $v \in S$ or $v$ is connected to $S$.
This implies that $S$ is a connected dominating set in $G$.

For the opposite direction, notice that any connected dominating set $S$ in $G$ implied a broadcast scheme for the data exchange problem.
If $v_0$ is in $S$, then consider a broadcasting scheme that transmits according to a Breadth First Search (BSF) starting from $v_0$ in the subgraph induced by $S$.
It is not hard to verify that such a scheme will use $|S|$ broadcasts and eventually will transmit $x_1$ to all the network.
Namely, let $(v_0,v_1,v_2\dots)$ be a BSF ordering from $v_0$ on the vertices of $S$.
The message $x_1 \in X$ can be transmitted from $v_0$ to all nodes in $V$ 
using the ordering $(v_0,v_1,v_2\dots)$.
Specifically, our ordering implies that node $v_j$ holds the message $x_1$ after nodes $\{v_0,v_1,\dots,v_{j-1}\}$ transmit and, as $S$ is dominating, all nodes 
will eventually receive the message $x_1$.
If $v_0$ is not is $S$ then it is connected to $s \in S$, so we can add $v_0$ to $S$ and still have a connected dominating set (and now use the scheme described in the last paragraph).
All in all, the resulting scheme will have $CDS + 1$ broadcasts.

As it is NP-hard to approximate $CDS$ within a multiplicative factor of $c\log{n}$ for some universal constant $c>0$ on graphs of size $n$ for which $CDS$ depends on $n$ (this follows directly from \cite{rs97} and Lemma~\ref{lem:CDS-DS^+}); it holds that the same is true for the parameter $NC$ under study.
\EPF

Note that the proof of Lemma~\ref{lem:k1} is also valid for $k=1$ in the specific case when only one vertex holds the information.
This implies that our upper bound for the case of disjoint sets of messages discussed in Section\ref{sec:disjoint} is tight.

\begin{figure}[t]
\begin{center}
 \includegraphics[scale=0.3, viewport=0 114 734 527]{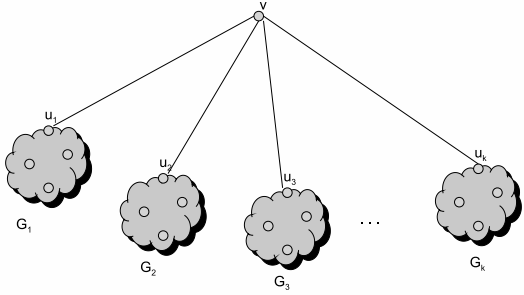}
\end{center}
\caption {Illustration of the graph $G'$ of Lemma~\ref{lem:hardness}.} \label{fig:NC-hardness}
\end{figure}

We now show that our hardness result holds for every $k$ by (again) presenting a reduction from the dominating set problem.
Given an instance $G=(V,E)$ to the dominating set problem, we construct the following graph $G'=(V',E')$ for the coded cooperative data exchange problem. $G'$ has $k$ copies of $G$, and a new vertex $v$, such that $v$ is connected to a vertex $u_i$ in each copy $G_i$ of $G$. Figure~\ref{fig:NC-hardness} illustrates $G'$. All vertices $u_i$ know all messages, $v$ knows no message, and for each $G_i$ all vertices in $G_i$ besides $u_i$ know all messages besides the $i$'th one.

\BL\label{lem:hardness}
If $DS(G)\le \alpha$ then $NC(G')\le k\cdot \frac{4}{3}\cdot(\alpha+1)$, and if $DS(G)> \beta$ then $NC(G')> k\cdot \beta$.
\EL
\BPF
Assume that $DS(G)\le \alpha$. Then the following is a transmission algorithm in $k\cdot \frac{4}{3}\cdot(\alpha+1)$ communication rounds.
Let $U_i$ be a minimum connected dominating set of $G_i$. For all $1\le i\le k$, as only one message needs to be broadcasted throughout $G_i$, one may design a broadcast scheme to satisfy all nodes in $G_i$ based on the connected dominating set $U_i$ exactly as in the proof of Lemma~\ref{lem:k1}. It holds (via Lemma~\ref{lem:CDS-DS^+}) that \BEQN NC(G')&\le& \sum_{i=1}^k\left(|U_i|+1\right)\le \sum_{i=1}^k\frac{4}{3}\cdot DS(G_i)+1\nonumber\\& =& k\cdot\left(\frac{4}{3}\cdot DS(G)+1\right)\le k\cdot\frac{4}{3}\cdot(\alpha+1)\nonumber.\EEQN

Now assume that $NC(G')\le k\cdot \beta$. We first show that $CDS(G_i)\le NC(G_i)$. This follows by the proof of Lemma~\ref{lem:k1}, as any communication scheme in $G_i$ only needs to communicate a single message $x_i$ from $u_i$ to the vertices of $G_i$ (recall that all vertices in $G_i$ know all the messages in $X \setminus \{x_i\}$).
Now, it also holds that $DS(G_i) \le CDS(G_i)$ and that $NC(G_i)\le NC(G')/k$, thus we obtain $DS(G) = DS(G_i)\le NC(G_i)\le \beta$.
All in all, we now conclude that estimating $NC(G')$ within a multiplicative factor of $O(\log{n})$ will imply such an estimate for $DS(G)$.
%
\EPF

Lemma~\ref{lem:k1}, Lemma~\ref{lem:hardness} and the hardness of computing $DS$ specified in \cite{rs97} imply the following theorem:

\BT
\label{the:hard}
The coded cooperative data exchange problem is NP-hard to approximate within $c\log \vert V\vert$, for some $c > 0$, for every value of $k$ polynomial in $|V|$.
\ET

\section{Approximation Algorithm}
\label{sec:alg}
In this section we give an approximation algorithm for the coded data exchange problem and analyze its approximation ratio. In the first subsection we present the approximation algorithm. In the second subsection we analyze the quality of the algorithm on a number of graph families or initial packet allocations, and show that for these instances the approximation ratio of the given algorithm matches (or comes close to matching) the results given in the previous section. In the third subsection we extend our analysis to the general case.

\subsection{The Algorithm}
\label{sec:alg1}
The following lemma introduces an approximation algorithm for the cooperative data exchange problem.
\BL\label{lem:alg}
Given a connected dominating set $D$ of $G$ 
one can efficiently solve the cooperative data exchange problem in 
$k \cdot(|D|+1)$
communication rounds.
Specifically, 
$NC\le k\cdot (CDS +1)$.
\EL
\BPF
The proof follows that given in Lemma~\ref{lem:k1}.
Let $D$ be a connected dominating set in $G$. 
Let $s_i$ be an arbitrary node holding message $x_i$. Assume that 
$s_i$ is a node in $D$.
Let $(s_i,v_1,v_2\dots)$ be a BSF (Breadth First Search) ordering from $s_i$ on the vertices of $D$.
The message $x_i \in X$ can be transmitted from $s_i$ to all nodes in $V$ 
using the ordering $(s_i,v_1,v_2\dots)$.
Specifically, our ordering implies that node $v_j$ holds the message $x_i$ after nodes $\{s_i,v_1,\dots,v_{j-1}\}$ transmit and, as $D$ is dominating, all nodes 
 will eventually receive the message $x_i$.
All in all, transmission of the $k$ messages will take $k\cdot CDS$ 
communication rounds.
If $s_i$ is not in $D$, then an additional round of communication is required for each message
in order for it to reach the set $D$.
\EPF

Since the problem of finding a minimum connected dominating set is NP-hard, we need to show how to approximately find such a set (efficiently).
Roughly speaking, we will find a connected dominating set in our network $G$ by first solving the fractional dominating set problem, by then {\em rounding} the fractional solution to an integral one to obtain a standard dominating set of $G$ (see e.g., \cite{l05,j74,s96,c79}), and by finally modifying the dominating set to a connected one via Lemma~\ref{lem:cds}.
All in all, this (well studied) scheme will yield a connected dominating set $D$ of size at most $c \log{n} \cdot DS_f$ for some universal constant $c>0$.

Repeating the above more formally, given an instance $G$ to the cooperative data exchange problem on general topologies, one can efficiently perform the following algorithm:
\begin{enumerate}
\item Solve the fractional dominating set problem on $G$ to obtain a fractional solution $\{c^f_v\}$.
\item Change the fractional solution to an integral one $\{c_v\}$ corresponding to a dominating set $D$ (via, e.g., \cite{l05,j74,s96,c79}).
\item Using $D$, construct a connected dominating set $D'$ (Lemma \ref{lem:cds}) with $|D'|=O(|D|)$.
\item Broadcast the $k$ source messages according to the procedure specified in Lemma~\ref{lem:alg} in $O(k|D'|) \leq O(k\log{n} \cdot DS_f)$ communication rounds.
\end{enumerate}

The procedure above will yield a communication scheme with at most $O(k\log{n} \cdot DS_f)$ communication rounds.
To understand the quality of the algorithm, one must express the size $NC$ (or at least bound it from {\em below}) by an expression which can be easily compared with the bound $O(k\log{n} \cdot DS_f)$.
For example, consider an instance to the data exchange problem in which $d=\max_{v\in V}{d_v}<k$ (here, for all $v_i \in V$ $d_{v_i}=|X_i|$).
We have seen via Lemma~\ref{lem:augmented-lb} that $NC \geq (k-d) \cdot DS^+_f \geq (k-d) \cdot DS_f \geq DS_f$.
Thus, on these instances we obtain a solution to the data exchange problem that is within a multiplicative factor of $O(k\log{n})$ from the optimal solution.
It is also not hard to see (we do this implicitly in Section~\ref{sec:talg}) that even if $d=\max_{v\in V}{d_v}=k$ a slight variant to our algorithm yields a solution which is within a multiplicative factor of $O(k\log{n})$ from the optimal solution.
The next sections attempt to improve this ratio to better match the hardness results presented in Section~\ref{sec:intract}.
Specifically, we show that the factor of $k$ in the ratio $O(k\log{n})$ can be reduced or in cases removed.

\subsection{Disjoint Sets of Messages}
\label{sec:disjoint}
In this subsection we analyze our approximation algorithm for the case that for each two nodes $v,u$ it holds that $X_v\cap X_u=\emptyset$. Note that this includes the case where only one node holds all the information, and all other nodes have no information. Namely, for some $v\in V$, $X_v=X$, and for all $u\neq v$ $X_u=\emptyset$.
For this case we are able to improve over the lower bound presented in Lemma~\ref{lem:augmented-lb}.

%
%
%
%
%

\BL\label{lem:augmented-sc}
$NC \geq k \cdot DS_f$.
\EL
\BPF
We show that a solution to the Coded Cooperative Data Exchange problem induces a solution to the $k$ Dominating Set problem.
For every vertex $v \in V$ define $c_v$ to be the number of packets transmitted by $v$ during an optimal data exchange protocol.
It holds that for every vertex $v$ the sum of capacities $c_u$ of all $u\in N(v)\cup \{v\}$ is at least $k$.
This is true since each node $v$ must send at least $|X_v|$ packets (as no other node holds the packets in $X_v$ and they must eventually reach the entire network), and receive at least $k-|X_v|=|\bar{X}_v|$ packets. Therefore $\sum_{u\in N(v)\cup \{v\}}{c_u} = \sum_{u\in N(v)}{c_u}+ c_v \ge |\bar{X_v}| + |X_v| = k$. Thus $(k-DS)_f\le k-DS \le NC$.
Finally, by Lemma~\ref{lem:kds+} it holds that $k \cdot DS_f = (k-DS)_f$.
\EPF

As our algorithm gives a communication scheme with at most $O(k\log{n}\cdot DS_f)$ rounds we conclude:


\BT
If for every two nodes $v,u$ it holds that $X_v\cap X_u=\emptyset$, 
the cooperative data exchange problem on general topologies can be efficiently solved within an approximation ratio of $O(\log{n})$.
Moreover, in such cases it holds that
$$
k\cdot DS_f\le NC\le k\cdot \left(\frac{4}{3}\cdot DS+1\right) \leq O(k\log{n})DS_f.
$$
As our algorithm does not involve coding, this implies that the coding advantage is $O(\log n)$.
\ET

\subsection{Regular Graphs}
\label{sec:regular}
In this subsection we show that if the given graph is regular our approximation algorithm has a $(1+\bar d /(k-\bar d))\cdot O(\log n)$ approximation ratio.
As before, we start by giving a lower bound for $NC$ in this case.
Let $G$ be a $\Delta$ regular graph, and let $\bar{d}=\frac{1}{n}\sum d_v$, then it holds that
\BL\label{lem:augmented-lb-reg}
$(k-\bar d)DS_f\le NC$.
\EL
\BPF
Consider the optimal communication scheme for the data exchange problem.
Since every vertex $v$ must receive at least $k-d_v$ messages, the total number of {\em edge} transmissions over the network is at least $\sum_{v\in V}{k-d_v}$ (here we are counting a single broadcast over $r$ edges as $r$ ``edge''-transmissions). Since each broadcast may transmit over at most $\Delta$ vertices it follows that
{\small{
\[NC\ge \frac{\sum_{v\in V}{k-d_v}}{\Delta} = \frac{n(k-\bar d)}{\Delta}\ge (k-\bar d)DS^+_f\ge (k-\bar d)DS_f.\]
}}
For the second inequality, notice that one can obtain a fractional self dominating set by setting $c_v=\frac{1}{\Delta}$ for each $v \in V$.
This implies that $DS^+_f \leq	\frac{n}{\Delta}$.
The last inequality holds by definition of $DS_f$ and $DS^+_f$.
\EPF

The following theorem follows from the above lemma:
\BT\label{theo:bounds}
The cooperative data exchange problem on regular topologies has a $(1+\bar d/(k-\bar d))\cdot O(\log{n})$ approximation ratio.
Specifically,
$$
(k-\bar d)\cdot DS_f\le NC\le k\cdot \left(\frac{4}{3}\cdot DS +1\right) \leq O(k\log{n})DS_f.
$$
As our algorithm does not involve coding, this implies that the coding advantage is $O\left(\left(1+\frac{\bar d}{k-\bar d}\right)\log n\right)$.
\ET
\BPF
By Lemma~\ref{lem:augmented-lb-reg} we have that
\BEQN
&&(k-\bar d)\cdot DS_f\le  NC\nonumber.
\EEQN

All in all, we obtain a solution with cost \BEQN O(\log{n}) \cdot k\cdot DS_f  &=& O(\log{n})\cdot \frac{k}{k-\bar d}\cdot(k-\bar d)DS_f\nonumber\\ &\leq& O(\log{n})\cdot NC\cdot \frac{k}{k-\bar d} \nonumber\\&=& O(\log{n})\cdot NC\cdot (1+\bar d/(k-\bar d))\nonumber.\EEQN
\EPF

\subsection{General Case}
\label{sec:General Case}
In this subsection we analyze the quality of our approximation algorithm for any instance $G$.
We first give an an example that shows that our lower bound for regular graphs stated in Lemma~\ref{lem:augmented-lb-reg} of $(k-\bar{d})\cdot DS_f$ 
 does not hold for general graphs.

\subsubsection{Example, complementing  Lemma~\ref{lem:augmented-lb-reg}}

\begin{figure}[t]
\begin{center}
  \includegraphics[scale=0.4, viewport=45 36 590 521]{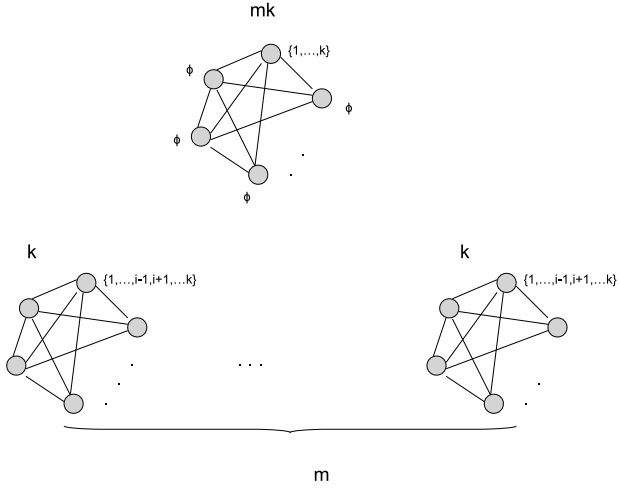}
\end{center}
\caption {Illustration of the graph $G$ for our {\em counter example} to Lemma~\ref{lem:augmented-lb-reg} on general (non-regular) graphs.} \label{fig:regular}
\end{figure}

We present a (general, non-regular) graph $G$ for which the lower bound stated in Lemma~\ref{lem:augmented-lb-reg} of $(k-\bar{d})\cdot DS_f$ does not hold (even in an approximate manner).
Consider a graph $G$ that consists of two parts:
The first part is a set of $m$ (disjoint) cliques of size $k$. In each
clique, for each message $i$ (between 1 to $k$)  there is exactly 1 vertex
missing message $i$ and having all the rest. The second part consists of a clique of size $mk$ in which one vertex
has all the information, and all the rest do not have any message.
Figure~\ref{fig:regular} illustrates $G$.
The value of an optimal scheme for data broadcast on the first part of $G$ is $2m$ since for each clique two messages must be sent (One client broadcasts an arbitrary message. This will cause another client to have all of $X$, and it broadcasts the sum of all the messages in $X$ over $F_q$). The value of an optimal scheme for data broadcast on the second part is obviously $k$ (just perform $k$ broadcasts from the single node that has all of $X$). Thus $NC$ is $2m+k$. Now, it is not hard to verify that $DS_f$ of the first part is $m$ (1 for each clique), and $DS_f$ of the second part is 1. Therefore $DS_f$ is $m+1$. Moreover, $\bar{d}=\frac{(k-1)km+mk}{2km}=\frac{k}{2}$, so $(k-\bar d)DS_f = \frac{k}{2}(m+1)$.
Therefore, for large $m\gg k$ we get that $NC\sim 2m$, and
$(k-\bar{d})DS_f\sim \frac{km}{2}$, so we get a gap of approximately $k/4$ w.r.t. the assertion of Lemma~\ref{lem:augmented-lb-reg}.

\subsubsection{Generalizing Lemma~\ref{lem:augmented-lb-reg}}
We use $\Delta$ to denote the maximum degree of $G$ and $\delta$ to denote the minimum degree of $G$.
We generalize Lemma~\ref{lem:augmented-lb-reg} to the case of general graphs:


\BL\label{lem:augmented-lb2}
$\frac{\delta}{\Delta}(k-\bar d)DS_f\le NC$.
\EL
\BPF
As in the proof of Lemma~\ref{lem:augmented-lb-reg}, 
the total number of edge transmissions over the network is at least $\sum_{v\in V}{k-d_v}$. Since each message can be transmitted to at most $\Delta$ vertices it follows that \BEQN NC&\ge& \frac{\sum_{v\in V}{k-d_v}}{\Delta} = \frac{n(k-\bar d)}{\Delta}\nonumber\\&\ge& \frac{\delta}{\Delta}(k-\bar d)DS^+_f\ge \frac{\delta}{\Delta}(k-\bar d)DS_f.\EEQN
In the setting at hand, the second inequality is valid since $\frac{n}{\delta}$ is an upper bound for $DS^+_f$ (i.e., one may set every $c_v$ to be equal to $\frac{1}{\delta}$ to get a valid solution to the linear program defining $DS^+_f$).
\EPF

We now conclude (recall that $d=\max_{v \in V}d_v$):

\BT\label{theo:bounds2}
The cooperative data exchange problem on general topologies has an approximation ration and coding advantage of
$$
O(\log n)\cdot \min\left\{\left(1+\frac{d}{k-d}\right),\frac{\Delta}{\delta}\left(1+\frac{\bar d}{k-\bar d}\right)\right\}.
$$ 
\ET
\BPF
Using Lemma \ref{lem:augmented-lb} we have that
\[(k-d)\cdot DS_f \le (k-d)\cdot DS^+_f\le NC.\]
In addition, by Lemma~\ref{lem:augmented-lb2} we have that
\BEQN
&&\frac{\delta}{\Delta}(k-\bar d)\cdot DS_f\le  NC,
\EEQN
Thus, the cost $O(\log{n}) \cdot k\cdot DS_f$ of our solution is at most:
$$
O(\log{n})\cdot NC\cdot \min\{(1+d/(k-d)),\frac{\Delta}{\delta}\cdot(1+\bar d/(k-\bar d))\}.
$$
\EPF


\subsection{A Tighter Upper Bound}
\label{sec:talg}
We now present a refined version of our algorithm from Section~\ref{sec:alg1}.
The algorithm we present will not yield improved asymptotic (in $n$) approximation ratios, however it yields improved communication schemes that at times may match those returned by the algorithm of Section~\ref{sec:alg1} and at times may be significantly better (depending on the instance at hand).

Roughly speaking, we improve the previous algorithm by taking into account the simple fact that it suffices to send each packet $x_i \in X$ only to those clients that do not hold it. Therefore we do not actually need to find a connected dominating set. Instead, we can do the following. 
Let $V_i$ be the set of vertices holding information packet $x_i$.
Let $\bar{V}_i = V \setminus V_i$.
A minimum sized $\bar V_i$-self dominating set 
is a minimum sized set of vertices $S \subset V$ such that each vertex in $\bar V_i$ has a neighbor in $S$, and each connected component of $S$ intersects $V_i$. 
Using $\bar V_i$-self dominating sets we can refine our algorithm.

Assume first that we know how to find a minimum sized $\bar V_i$-self dominating set 
for each message $x_i$. 
Let $\{C^i_j\}_{j=1}^{\ell}$ be the set of connected components of such a minimum sized $\bar V_i$-self dominating set. 
Let $w_j$ be an arbitrary vertex in $C^i_j\cap V_i$.
To communicate $x_i$ throughout $C^i_j$ we use the following natural procedure: $w_j$ sends $x_i$, and then each vertex in $C^i_j$ that received $x_i$ sends $x_i$. It immediately follows, after performing this process for each connected component $C^i_j$, that all vertices in $G$ hold $x_i$. Moreover, the number of communication rounds used in this scheme is equal to the size of the $\bar V_i$-self dominating set. Let $DS_i=DS_i(G)$ denote the minimum size of a $\bar V_i$-self dominating set in $G$.

\begin{figure}[t]
\begin{center}
 \includegraphics[scale=0.3, viewport=62 19 489 520]{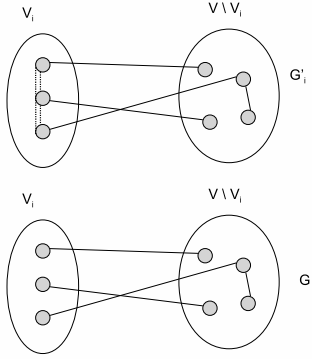}
 \end{center}
\caption {Illustration of $G'_i$ to be used in the construction of $DS_i$.} \label{fig:DS_i-reduction}
\end{figure}

We now turn to approximating $DS_i$. 
We define the following graph $G'_i=(V_i',E_i')$ corresponding to our definition of a $\bar V_i$-self dominating set: 
$V_i' = V_i$, and $E_i' = E\cup (V_i\times V_i)$.
Figure~\ref{fig:DS_i-reduction} illustrates the construction of $G'_i$.

\BL\label{lem:DS_i}
$CDS(G_i') \leq DS_i(G) \leq CDS(G_i') +1$.
\EL
\BPF
Let $D$ be a minimum sized $\bar V_i$-self dominating set in $G$. Then by definition of $DS_i$ it follows that $D$ is a connected dominating set in $G'_i$, since every connected component of $D$ in $G$ has a vertex in $V_i$, and all vertices in $V_i$ are connected in $G'_i$. 
Similarly, let $D$ be a minimum connected dominating set in $G'_i$.
If $D$ includes a vertex in $V_i$ then it follows that $D$ is also a $\bar V_i$-self dominating set in $G$.
Otherwise (as we assume w.l.o.g. that $G$ is connected) we add one vertex $v$ in $V_i$ to $D$. Here, we take $v$ to be any vertex in $V_i$, as they are all connected to $D$.
This completes the proof.\EPF

By Lemma~\ref{lem:DS_i} we can efficiently perform the following algorithm:
\BE
\item For $1\le i \le k$ do:
\BE
\item Construct $G_i'$. 
\item Using the algorithm specified in Section~\ref{sec:alg} 
construct a connected dominating set $D_i$ in $G_i'$ via the corresponding fractional solutions, with $|D_i|=O(\log|V'_i|\cdot DS_f(G'_i))$.
\item For each connected component of $D_i$ in $G$ 
broadcast $x_i$ (according to the procedure specified in the discussion above) in $O(|D_i|)$ communication rounds.
\EE
\EE

All in all, the refined algorithm efficiently solves the data exchange problem in $O(\sum_i \log|V'_i|\cdot DS_f(G'_i))$ rounds of communication which is at most the number $O(k\log n\cdot DS_f(G))$ of rounds from the original algorithm.
This follows since $G$ is  subgraph (in edges) of $G'_i$, and thus by definition $DS_f(G'_i) \leq DS_f(G)$. Thus, our refined algorithm is at least as good as that of Section~\ref{sec:alg1} and improves over it in cases in which $DS_i$ is significantly smaller than the dominating set in $G$.

\section{Concluding remarks}
\label{sec:diss}
In this paper, we consider the cooperative
data exchange problem for general topologies $G$ in the combinatorial {\em integral} setting.
We establish both upper and lower bounds on the
multiplicative approximation ratio that one may obtain efficiently by tying our problem to certain well studied combinatorial properties of $G$.
Our achievability results are based on communication schemes that do not involve coding and in such imply bounds on the coding advantage of the problem at hand.
Our results address the setting of undirected networks.
Extending our results to the case of directed graphs (by studying directed analogs to dominating sets) involves modifications in our analysis and is subject to future research.

 \bibliographystyle{plain}
\bibliography{CodedCooperative}

\end{document}